\documentclass[preprint,showpacs,preprintnumbers,amsmath,amssymb]{revtex4}

\usepackage{graphicx}
\usepackage{epsfig}		
\usepackage{dcolumn}
\usepackage{bm}

\def\0{\mbox{\tiny $0$}}
\def\1{\mbox{\tiny $1$}}
\def\2{\mbox{\tiny $2$}}
\def\3{\mbox{\tiny $3$}}
\def\4{\mbox{\tiny $4$}}
\def\5{\mbox{\tiny $5$}}
\def\6{\mbox{\tiny $6$}}
\def\7{\mbox{\tiny $7$}}
\def\8{\mbox{\tiny $8$}}
\def\9{\mbox{\tiny $9$}}
\def\N{\mbox{\tiny $N$}}
\def\R{\mbox{\tiny $R$}}

\def\bb#1{\mbox{\footnotesize $(#1)$}}

\def\x{\mbox{\tiny $x$}}

\def\L{\mbox{\tiny $L$}}
\def\ii{\mbox{\tiny $i$}}

\def\D{\mbox{\tiny $D$}}

\def\pmp{\mbox{\tiny $\mp$}}

\begin{document}
\title{Delay time computation for relativistic tunneling particles}

\author{A. E. Bernardini}
\email{alexeb@ifi.unicamp.br}
\affiliation{Instituto de F\'{\i}sica Gleb Wataghin, UNICAMP,
PO Box 6165, 13083-970, Campinas, SP, Brasil.}
\altaffiliation[Also at] {~IST, Departamento de F\'{\i}sica, Av. Rovisco Pais, 1, 1049-001, Lisboa, Portugal.}

\date{\today}

\begin{abstract}
We study the tunneling zone solutions of a one-dimensional electrostatic potential for the relativistic (Dirac to Klein-Gordon) wave equation when the incoming wave packet exhibits the possibility of being almost totally transmitted through the barrier.
The transmission probabilities, the phase times and the dwell times for the proposed relativistic dynamics are obtained and the conditions for the occurrence of accelerated tunneling transmission are all quantified.
We show that, in some limiting cases, the analytical difficulties that arise when the stationary phase method is employed for obtaining phase (traversal) tunneling times are all overcome.
Lessons concerning the phenomenology of the relativistic tunneling suggest revealing insights into condensed-matter experiments using electrostatic barriers for which the accelerated tunneling effect can be observed.
\end{abstract}

\date{\today}
\maketitle

\section{Introduction}

Finding a definitive interpretation for the Nature of {\em superluminal} barrier tunneling has brought up a fruitful discussion in the literature \cite{But03,Win03A,Win03,Olk04} since pulses of light and microwaves appear to tunnel through a barrier at speeds faster than a reference pulse moves through a vacuum \cite{Nim92,Ste93,Spi94,Nim94,Hay01}.
Tunneling occurs when a wave impinges on a thin barrier of opaque material and some small amount of the wave {\em leaks} through to the other side.
The superluminal experiments that promoted the controversial discussions were performed with a lattice of layers of transparent and opaque materials arranged so that waves of some frequencies are reflected (through destructive interference) but other frequencies pass through the lattices in a kind of {\em filter} effect correlated to the Hartman effect \cite{Har62}.

To obtain a definitive answer for the time spent by particle to penetrate a classically forbidden region delimited by a potential barrier, people have tried to introduce quantities that have the dimension of time and can somehow be associated with the passage of the particle through the barrier or, strictly speaking, with the definition of the tunneling time.
These proposals have led to the introduction of several {\em transit time} definitions that can be summarized by three groups.
(1) The first group comprises a time-dependent description in terms of wave packets where some features of an incident packet and the comparable features of the transmitted packet are utilized to describe a quantifiable {\em delay} as a tunneling time \cite{Hau89}.
(2) In the second group the tunneling times are computed based on averages over a set of kinematical paths, whose distribution is supposed to describe the particle motion inside a barrier.
In this case, Feynman paths are used like real paths to calculate an average tunneling time with the weighting function $\exp{[i\, S\, x(t)/\hbar]}$, where $S$ is the action associated with the path $x(t)$ (where $x(t)$ represents the Feynman paths initiated from a point on the left of the barrier and ending at another point on the right of it \cite{Sok87}).
The Wigner distribution paths \cite{Bro94}, and the Bohm approach \cite{Ima97,Abo00} are included in this group.
(3) In the third group we notice the introduction of a new degree of freedom, constituting a physical clock for the measurements of tunneling times.
This group comprises the methods with a Larmor clock \cite{But83} or an oscillating barrier \cite{But82}.
Separately, standing on itself is the {\em dwell} time defined by the interval during which the incident flux has to exist and act, to provide the expected accumulated particle storage, inside the barrier \cite{Lan94}.
In spite of no general agreement \cite{Olk04,Olk92} among the above definitions, the so called phase time \cite{Wig55} (group delay) and the dwell time have an apparently well established relation between them \cite{Hau89,Win03}.
However, these time definitions remain controversial since in the opaque barrier limit they predict effective tunneling velocities that exceed the vacuum speed of light and may even become unlimited (Hartman effect)\cite{Har62}.

In this scenario, some of the barrier traversal time definitions lead, under tunneling time conditions, to very short times, which can even become negative.
It can precipitately induces an interpretation of violation of simple concepts of causality.
Otherwise, negative speeds do not seem to create problems with causality, since they were predicted both within special relativity and within quantum mechanics \cite{Olk95}.
A possible explanation for such time advancements can come, in any case, from consideration of the very rapid spreading of the initial and transmitted wave packets for large momentum distribution widths.
Due to the similarities between tunneling (quantum) packets and evanescent (classical) waves, exactly the same phenomena are to be expected in the case of classical barriers\footnote{In particular, we could mention the analogy between the stationary Helmholtz equation for an electromagnetic wave packet - in a waveguide, for instance - in the presence of a {\em classical} barrier and the stationary Schroedinger equation, in the presence of a potential barrier \cite{Nim94,Lan94,Jak98}).}.
The existence of such negative times is predicted by relativity itself based on its ordinary postulates \cite{Olk04}, and they appear to have been experimentally detected in many works \cite{Gar70,Chu82}.

In all cases described by the non-relativistic (Schroedinger) dynamics \cite{Olk04}, the pulse (wave packet) that emerges from the tunneling process is greatly attenuated and front-loaded due to the {\em filter} effect (only the leading edge of the incident wave packet survives the tunneling process without being severally attenuated to the point that it cannot be detected).
If one measures the speed by the peak of the pulse, it looks faster than the incident wave packet.
Since the transmission probability depends analytically on the momentum component $k$  ($T \equiv T\bb{k}$) the initial (incident wave) momentum distribution can be completely distorted by the presence of the barrier of potential.
As there is no sharp beginning to a pulse, we cannot determine the instant of its arrival at a certain point.
Thus the computation of the tunneling time becomes fundamentally meaningless.
We could only watch the rising edge of the pulse and try to recognize what is arriving.

By observing the tunneling dynamics described by the relativistic Dirac/Klein-Gordon equation, we can overcome all of these difficulties.
We demonstrate with complete mathematical accuracy that, in some limiting cases of the
relativistic (Klein) tunneling phenomena where the relativistic kinetic energy is approximately equal to the potential energy of the barrier, and $m c L /\hbar << 1$, particles with mass $m$ can pass
through a potential barrier $V_{\0}$ of width $L$ with transmission probability $T$ approximately equal to one.
Since $T \sim 1$, the analytical conditions for the stationary phase principle applicability which determines the tunneling (phase) time for the transmitted wave packets are totally recovered.
Differently from the previous (non-relativistic) tunneling analysis, the original momentum is kept undistorted and there is no {\em filter} effect.
We shall demonstrate that the tunneling time is then computed for a completely undistorted transmitted wave packet, which legitimizes any eventual accelerated transmission.

In this manuscript, we perform the calculations to obtain the tunneling transmission probabilities and the
respective delay times of the corresponding wave packets when a relativistic dynamics is taken into account.
In section II, we introduce the relativistic dynamics that circumvents the tunneling phenomenon and we define some novel parameters which will be useful for the subsequent analysis.
The transmission probabilities, the phase times and the dwell times for the relativistic dynamics that we have proposed are obtained in section III.
We also quantify the conditions for the occurrence of accelerated and, eventually, superluminal tunneling transmission probabilities.
We draw our conclusions in section IV.

\section{Defining the dynamics, variables and limits}

It is very difficult and probably even confusing to treat all interactions of plane waves or wave packets with a barrier potential using a relativistic wave equation \cite{Del03,Cal99,Dom99,Che02}.
This is because the physical content depends upon the relation between the barrier height $V_{\0}$ and the mass $m$ of the incoming (particle) wave, beside of its total energy $E$.
In the first attempt to evaluate this problem, Klein \cite{Kle29} considered the reflection and transmission of electrons of  incidence energy $E$ on the potential step $V\bb{x} = \Theta\bb{x}V_{\0}$ in the $(2+1)$-dimensional time-independent Dirac equation which can be represented in terms of the usual Pauli matrices \cite{Zub80} by\footnote{$\Theta\bb{x}$ is the Heavyside function.}
\small\begin{equation}
\left[\sigma^{3}\sigma^{\ii}\partial_{\ii} - (E - \Theta\bb{x_{\1}}V_{\0}) - \sigma^{3} m\right]\phi\bb{k, x_{\1},x_{\2}} = 0,
~~(\mbox{from this point}~ c = \hbar = 1),
\label{001}
\end{equation}\normalsize
which corresponds to the reduced representation of the usual Pauli-Dirac {\em gamma} matrix representation ($i = 1,\, 2$).
The physical essence of such a theoretical configuration lies in the prediction that fermions can pass through large repulsive potentials without exponential damping.
It corresponds to the so called {\em Klein tunneling} phenomenon \cite{Cal99} which follows accompanied by the production of a particle-antiparticle pair inside the potential barrier.
It is different from the usual tunneling effect since it occurs in the energy zone of the Klein paradox \cite{Kle29,Zub80}.
Taking the quadratic form of the above equation reduced to $(1+1)$-dimension for a generic scalar potential $V\bb{x}$, we obtain the analogous Klein-Gordon equation,
\small\begin{equation}
\left(i \partial_{\0} - V\bb{x}\right)^{\2}\phi\bb{k, x} = \left(-\partial^{\2}_{\x} + m^{\2}\right)\phi\bb{k, x},
\label{002}
\end{equation}\normalsize
which, from the mathematical point of view, due to the second-order spatial derivatives, has similar boundary conditions to those ones of the Schroedinger equation and leads to stationary wave solutions characterized by a {\em relativistically} modified dispersion relation.

By depicting three potential regions by means of a rectangular potential barrier $V\bb{x}$, $V\bb{x} = V_{\0}$ if $0 \leq x \leq L$, and $V\bb{x} = 0$ if $x < 0$ and $x > L$, differently from the non-relativistic (Schroedinger) dynamics, we observe that the incident energy can be divided into three zones.
The {\em above barrier} energy zone, $E > V_{\0} + m$, involves diffusion phenomena of oscillatory waves (particles).
In the so called {\em Klein} zone \cite{Kle29,Cal99}, $E < V_{\0} - m$, we find oscillatory solutions (particles and antiparticles) in the barrier region.
In this case, antiparticles see an opposite electrostatic potential to that seen by the particles and hence they will see a well potential where the particles see a barrier \cite{Aux1,Kre04}.
The {\em tunneling} zone, $V_{\0} - m < E < V_{\0} + m$, for which only evanescent waves exist \cite{Kre01,Pet03} in the barrier region, is that of interest in this work.
In fact, the usual definition of {\em Klein tunneling}, which involves the Klein paradox \cite{Kle29}, has been studied in the literature just for the Dirac equation \cite{Dom99,Cal99,Pet03,Tel95,Han99}.
In the following, we do not aim to treat the Klein paradox (not evanescent) ``tunneling'' zone, but the evanescent tunneling zone, several times ignored in the analysis of relativistic tunneling.
The point is that, to treat the evanescent tunneling zone, from the mathematical point of view which involves the equation of motion, it is sufficient to consider the quadratic form of the Dirac equation, namely the Klein-Gordon  equation.
For the evanescent tunneling zone, the Dirac equation and its quadratic form lead to the same results when we apply the (evanescent) tunneling time definitions in which we are interested, the phase time and the dwell time.
The evanescent tunneling zone does not intersect with the Klein paradox energy zone for which, at least theoretically, the possibility of creation/annihilation of fermionic pairs leads to the reinterpretation of the probability density currents, and thus to a novel interpretation of the tunneling phenomenon.
Consequently, concerning the calculation of evanescent tunneling times, all the references to fermionic (Dirac) and bosonic (Klein-Gordon) particles are valid, in the same sense that all the results derived from the non-relativistic Schr\"{o}dinger equation are supposed to be valid for massive fermions and massive bosons.
Being more specific, our principal focus here is the calculation of phase times and dwell times and the possibility of a Hartman-like effect \cite{Har62} independent of the barrier width.

By evaluating the problem for this tunneling (evanescent) zone assuming that $\phi (k,x)$ are stationary wave solutions of the Eq.~(\ref{002}), when the peak of an incident (positive energy) wave packet reach the barrier $x = 0$ at $t = 0$, we can usually write
\small\begin{equation}
\phi(k,x)=
\left\{\begin{array}{l l l l}
\phi_{\1}(k,x) &=&
\exp{\left[ i \,k \,x\right]} + R(k,L)\exp{\left[ - i \,k \,x \right]}&~~~~x < 0,\nonumber\\
\phi_{\2}(k,x) &=& \alpha(k)\exp{\left[ - \rho\bb{k}  \,x\right]} + \beta(k)\exp{\left[ \rho\bb{k}  \,x\right]}&~~~~0 < x < L,\nonumber\\
\phi_{\3}(k,x) &=& T(k,L)\exp{ \left[i \,k (x - L)\right]}&~~~~x > L,
\end{array}\right.
\label{003}
\end{equation}\normalsize
where the dispersion relations are modified with respect to the usual non-relativistic ones: $k^{\2} = E^{\2} - m^{\2}$ and $ \rho\bb{k}^{\2} = m^{\2} - (E - V_{\0})^{\2}$.

In order to proceed with a phenomenological analysis which allows us to establish a correspondence with the non-relativistic (NR) solutions, it is convenient to define the kinematic variables in terms of the following parameters: $w = \sqrt{2 m V_{\0}}$, $\upsilon = V_{\0}/m = w^{\2}/2m^{\2}$, and $n^{\2}\bb{k} = k^{\2}/w^{\2} = E_{\N\R}/V_{\0}$.
The parameter $w$ corresponds to the same {\em normalizing} parameter of usual NR analysis where $k^{\2} = 2 m E_{\N\R}$.
The previously quoted relation between the potential energy $V_{\0}$ and the mass $m$ of the incident particle is given by the parameter $\upsilon$.
Finally, $n^{\2}\bb{k}$ represents the dependence on the energy for all the results that will be considered here.
Let us then remind that the energies for which the evanescent tunneling occurs are comprised by the above mentioned interval $V_{\0} - m < E < V_{\0} + m$.
By adding $m$ or $-m$ to each term of this interval, we obtain
\small\begin{equation}
V_{\0} - m \leq E \leq V_{\0} + m ~~\Rightarrow \left\{\begin{array}{ll}V_{\0} \leq E + m \leq V_{\0} + 2 m \\
V_{\0} - 2 m \leq E - m \leq V_{\0} \end{array}\right.
\label{003B}
\end{equation}\normalsize
Since $E$, $V_{\0}$ and $m$ are all positive quantities, we can multiply the terms the first interval expression by the second one, in order to obtain
\small\begin{equation}
V_{\0}(V_{\0} - 2 m) \leq E^{\2} - m^{\2} \leq V_{\0}(V_{\0} + 2m).
\label{003C}
\end{equation}\normalsize
By observing that $E^{\2} - m^{\2} = k^{\2}$, subtracting $V_{\0}^{\2}$ from the three terms and dividing the resultant terms by $2 m V_{\0}$, we get the final inequality
\small\begin{equation}
-1 \leq \frac{k^{\2}}{2 m V_{\0}} - \frac{V_{\0}}{2 m} \leq 1.
\label{003D}
\end{equation}\normalsize
which can be squared in order to set that the tunneling zone for the above form of the Klein-Gordon equation (\ref{002}) is comprised by the interval $(n^{\2}\bb{k} - \upsilon/2)^{\2} \leq 1$ allowing that $n^{\2}\bb{k}$ might assume larger values ($n^{\2}\bb{k} >> 1$), in opposition to the NR case where the tunneling energy zone is constrained by $0 < n^{\2}\bb{k} < 1$).
We shall observe that such a peculiarity has a subtle relation with the possibility of superluminal transmission through the barrier.
The limits for NR energies ($k^{\2} << m^{\2}$ and $V << m$) are given by $\upsilon n^{\2} << 1$ and $\upsilon/n^{\2} << 1$, which as we shall notice in the next section reproduces the transmission and delay results of the Schroedinger equation.

\section{The transmission coefficient and the delay times}

The stationary phase method can be successfully applied for describing the movement of the center of a wave packet constructed in terms of a symmetrical momentum distribution $g(k - k_{\0})$ which has a pronounced peak around $k_{\0}$.
By assuming that the phase that characterizes the propagation varies smoothly around the maximum of $g(k - k_{\0})$, the stationary phase condition enables us to calculate the position of the peak of the wave packet (highest probability region to find the propagating particle).
With regard to the {\em standard} one-way direction wave packet tunneling, for the set of stationary wave solutions given by Eq.~(\ref{003}), it is well-known \cite{Ber06} that the transmitted amplitude $T\bb{n, L} = |T\bb{n, L}|\exp{[i \varphi\bb{n, L}]}$ is written in terms of
\small\begin{equation}
|T\bb{n, L}| = \left\{1 + \frac{1}{4 \, n^{\2} \, \rho^{\2}\bb{n}} \sinh^{\2}{\left[\rho\bb{n}\, w L \right]}\right\}^{-\frac{1}{2}},
\label{004}
\end{equation}\normalsize
where we have suppressed from the notation the dependence on $k$,
and
\small\begin{equation}
\varphi\bb{n, L} = \arctan{\left\{\frac{n^{\2} - \rho^{\2}\bb{n}}
{2 n \, \rho\bb{n}}
\tanh{\left[\rho\bb{n} \, w L \right]}\right\}},
\label{005}
\end{equation}\normalsize
for which we have made explicit the dependence on the barrier length $L$ (parameter $w L$)
and we have rewritten $\rho\bb{k} = w \rho\bb{n}$, with $\rho\bb{n}^{\2} = \sqrt{1 + 2 n^{\2} \upsilon} - (n^{\2} -\upsilon/2)$.

\begin{figure}[th]
\epsfig{file=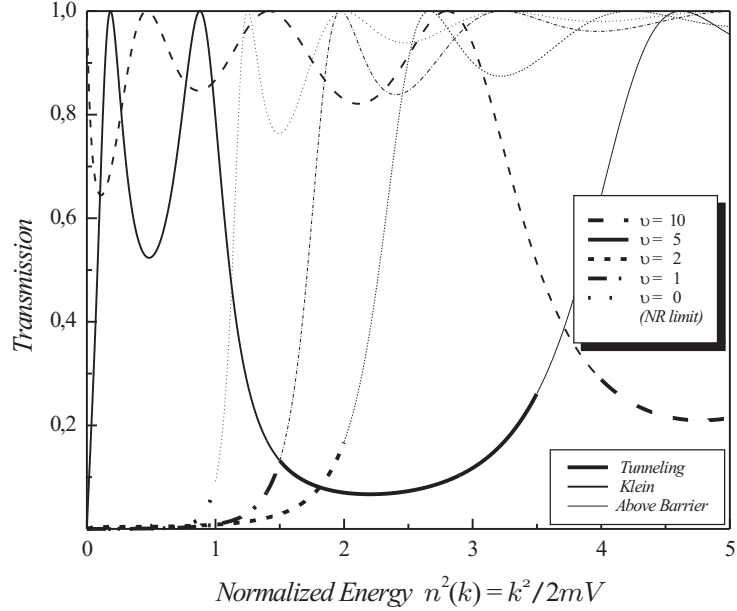,width=11cm}
\caption{Tunneling TRANSMISSION probabilities $|T\bb{n, L}|^{\2}$ for the incidence on a rectangular potential barrier of height $V_{\0}$ for the dynamics of a relativistic wave equation versus the normalized energy $n^{\2}\bb{k}$.
We have classified the energy zones by the line thickness: the thick line corresponds to the tunneling zone, the intermediate line corresponds to the Klein zone and the thin line corresponds to the above barrier zone.
Here we have adopted the illustratively convenient value of $w L = 2 \pi$ and have set $\upsilon =  0,\, 1,\, 2 ,\, 5,\, 10$, where NR regime can be parameterized by $\upsilon = 0$.
We have constrained our analysis to $n^{\2}\bb{k} > 0$ since we have assumed $V_{\0} > 0$.}
\label{Fig01}
\end{figure}

We illustrate the modified tunneling transmission probabilities in the Fig.~\ref{Fig01} for different propagation regimes ($\upsilon = 0 (NR),\, 1,\, 2,\,5,\, 10)$)
by observing that the tunneling region is comprised by the interval $(n - \upsilon/2)^{\2} < 1,\, n > 0$.

The additional phase $\varphi(n, L)$ that goes with the transmitted wave is utilized for calculating the transit time $t_{\varphi}$ of a transmitted wave packet when its peak emerges at $x = L$,
\small\begin{equation}
t_{\varphi} = \frac{\mbox{d}k}{\mbox{d}E\bb{k}} \frac{\mbox{d}n\bb{k}}{\mbox{d}k} \frac{\mbox{d}\varphi\bb{n, L}}{\mbox{d}n} = \frac{(L)}{v} \frac{1}{w (L)} \frac{\mbox{d}\varphi\bb{n, L}}{\mbox{d}n},
\label{006}
\end{equation}\normalsize
evaluated at $k = k_{\0}$ (the maximum of a generic symmetrical momentum distribution $g(k - k_{\0})$ that composes the {\em incident} wave packet).
By introducing the {\em classical} traversal time defined as
$\tau_{\bb{k}} = L (\mbox{d}k/\mbox{d}E\bb{k})= L / v$,
we can obtain the normalized phase time,
\small\begin{equation}
\frac{t_{\varphi}}{\tau_{\bb{k}}}
 = \frac{f\bb{n, L}}{g\bb{n, L}}
\label{007}
\end{equation}\normalsize
where
\small\begin{eqnarray}
f\bb{n, L} &=&
8 n^{\2} \left[\left(2 + 8 n^{\2} \upsilon + \upsilon^{\2}\right) - \left(4 n^{\2} + 3 \upsilon\right)\sqrt{1 + 2 n^{\2}\upsilon} \right]\nonumber\\
&&~~~~+
4 \left[\left(4 + 4 n^{\2}\upsilon + \upsilon^{\2}\right)\sqrt{1 + 2 n^{\2}\upsilon} - 2 \upsilon \left(2 + 3 n^{\2}\upsilon\right)\right]\frac{Sh(\rho\bb{n} w L)\,Ch(\rho\bb{n} w L)}{\rho\bb{n} w L}
\nonumber
\end{eqnarray}\normalsize
and
\small\begin{eqnarray}
g\bb{n, L} &=&
16 n^{\2} \left[2 \left(1 + 2 n^{\2}\upsilon\right) -  \sqrt{1 + 2 n^{\2}\upsilon}\left(2 n^{\2} + \upsilon\right)\right]\nonumber\\
&&~~~~+
2 \left[\left(4 + 8 n^{\2}\upsilon + \upsilon^{\2}\right)\sqrt{1 + 2 n^{\2}\upsilon} - 4 \upsilon \left(1 + 2 n^{\2}\upsilon\right)\right]Sh(\rho\bb{n} w L)^{\2}
\nonumber
\end{eqnarray}\normalsize
where $Ch(x) = \cosh{(x)}$ and $Sh(x) = \sinh{(x)}$.

To illustrate the above results we plot the tunneling phase times in correspondence with the
transmission probabilities of the Fig.~\ref{Fig01} for the same different propagation regimes ($\upsilon = 0 (NR),\, 1,\, 2,\,5,\,10)$).
\begin{figure}[th]
\epsfig{file=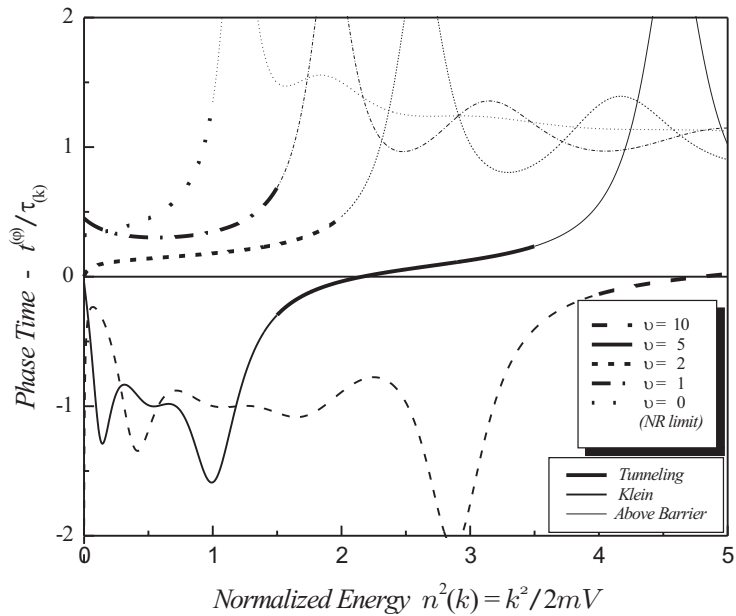,width=11cm}
\caption{The related tunneling PHASE times in correspondence with their the tunneling TRANSMISSION probability for the dynamics of the Klein-Gordon form equation.
Again by means of the line thickness we observe that the tunneling region is comprised by the interval $(n^{\2} - \upsilon/2)^{\2} < 1,\, n > 0$.
We have used the same criteria and the same set of parameters from the Fig.~\ref{Fig01}.}
\label{Fig02}
\end{figure}

By observing the results illustrated by the above figures, we can notice the possibility of accelerated ($t_{\varphi} < \tau_{\bb{k}}$), and eventually {\em superluminal} (negative tunneling delays, $t_{\varphi} < 0$) transmissions without recurring to the usual analysis of the {\em opaque} limit ($\rho\bb{n} w L \rightarrow \infty$) which leads to the Hartman effect \cite{Har62}.
In the NR dynamics (Schroedinger equation solutions), the opaque limit and the consequent superluminal interpretation of the results of such an approximation (Hartman effect) were extensively discussed in the literature.
Superluminal group velocities in connection with quantum (and classical) tunneling were predicted even on the basis of tunneling time definitions more general than the simple Wigner's phase time \cite{Wig55} (Olkhovsky {\em et al.}, for instance, discuss a simple way of understanding the problem \cite{Olk04}).
Experiments with tunneling photons and evanescent electromagnetic waves \cite{Nim92,Ste93,Hay01} have generated a lot of discussions on relativistic causality, which,
in addition to several analytical limitations, have ruined some possibilities of superluminal interpretation of the tunneling phenomena \cite{Lan89,Win03,Ber06}.
In a {\em causal} manner, the above arguments might consist in explaining the superluminal phenomena during tunneling as simply due to a {\em reshaping} of the pulse, with attenuation, as already attempted (at the classical limit) \cite{Gav84}, i. e. the later parts of an incoming pulse are preferentially attenuated, in such a way that the outcoming peak appears shifted towards earlier times even if it is nothing but a portion of the incident pulse forward tail \cite{Ste93,Lan89}.

The Hartman effect is related to the fact that for opaque potential barriers the mean tunneling time does not depend on the barrier width, so that for large barriers the effective tunneling-velocity can become arbitrarily large, where it was found that the tunneling phase time was independent of the barrier width.
It seems that the penetration time, needed to cross a portion of a barrier, in the case of a very long barrier starts to increase again after the plateau corresponding to infinite speed — proportionally to the distance \footnote{The validity of the HE was tested for all the other theoretical expressions proposed for the mean tunneling times \cite{Olk04}.}.

We do not intend to extend on the delicate question of whether superluminal group-velocities can sometimes imply superluminal signalling, a controversial subject which has been extensively explored in the literature (\cite{Olk04} and references therein).
Otherwise, the phase time calculation based on the relativistic dynamics introduced here offers distinct theoretical possibilities in a novel scenario, for the limit case where $\rho\bb{n}\,L$ tends to $0$ (with $L \neq 0$), in opposition to the opaque limit where $\rho\bb{n}$ tends to $\infty$.
Let us then separately expand the numerator $f\bb{n, L}$ and the denominator $g\bb{n, L}$ of the Eq.~(\ref{007}) in a power series of $\rho\bb{n}\,wL$ ($\rho\bb{n}  \rightarrow 0$) in order to observe that in the lower (upper) limit of the tunneling energy zone, where
$n^{\2}$ tends to $\upsilon/2 + (-) 1$, the numerical coefficient of the zero order term in $\rho\bb{n}\,wL$ amazingly vanish in the numerator as well as in the denominator!
Since the coefficient of the linear term also is null,
just the coefficient of the second order terms plays a relevant role in both series expansions.
After expanding the Eq.~(\ref{007}), such a {\em step-by-step} mathematical exercise leads to
\small\begin{equation}
\frac{t_{\varphi}}{\tau_{\bb{k}}}
 = \frac{4}{3}
\frac{\left[\left(4 + 4 n^{\2}\upsilon + \upsilon^{\2}\right)\sqrt{1 + 2 n^{\2}\upsilon} - 2 \upsilon \left(2 + 3 n^{\2}\upsilon\right)\right]}{\left[\left(4 + 8 n^{\2}\upsilon + \upsilon^{\2}\right)\sqrt{1 + 2 n^{\2}\upsilon} - 4 \upsilon \left(1 + 2 n^{\2}\upsilon\right)\right]}
+\mathcal{O}(\rho\bb{n}\,wL)^{\2}
\label{0100}
\end{equation}\normalsize
for small values of $\rho\bb{n}$.
At the same time, since $\lim_{n^{\2}\rightarrow \upsilon/2 \pmp 1}{\rho\bb{n}} = 0$, the tunneling transmission probability can be approximated by
\small\begin{equation}
\lim_{n^{\2}\rightarrow \upsilon/2 \pmp 1}{\rho\bb{n}} = 0.
\label{010}
\end{equation}\normalsize
By taking the above limit, we easily obtain,
\small\begin{equation}
\lim_{n^{\2}\rightarrow \upsilon/2 \pmp 1}{|T\bb{n, L}|} =
\left[1 + \frac{(w L)^{\2}}{2 \upsilon \mp 4}\right]^{-\frac{1}{2}}
\begin{array}{c}\mbox{\tiny$\upsilon >> 1$}\\ \rightarrow \\~\end{array}
\left[1 + (m L)^{\2}\right]^{-\frac{1}{2}},
\label{011}
\end{equation}\normalsize
from which we recover the possibility of a highly probable tunneling transmission when $m L << 1$.
Finally, for the corresponding values of the phase times evaluated in (\ref{0100}), we obtain,
\small\begin{equation}
\lim_{n^{\2}\rightarrow \upsilon/2 \mp 1}{\frac{t_{\varphi}}{\tau_{\bb{k}}}} = -\frac{4}{3}\frac{1}{1 \pm 2 n^{\2}}, ~~~~ n^{\2}\rightarrow \upsilon/2 \mp 1, ~~~n^{\2},\,\upsilon > 0,
\label{012}
\end{equation}\normalsize
that does not depend on $m L$, and we notice that its asymptotic limit always converges to $0$.
In particular, in the lower limit of the tunneling energy zone, $n^{\2}\rightarrow \upsilon/2 - 1$, it is always negative.
Since the result of Eq.~(\ref{012}) is exact, and we have accurately introduced the possibility of obtaining total transmission ({\em transparent barrier}) our result ratifies the possibility of accelerated transmission (positive time values), and consequently superluminal tunneling (negative time values), for relativistic particles when $m L$ is sufficiently smaller than 1 $(\Rightarrow T \simeq 1)$.

\begin{figure}
\vspace{-0.6 cm}
\epsfig{file=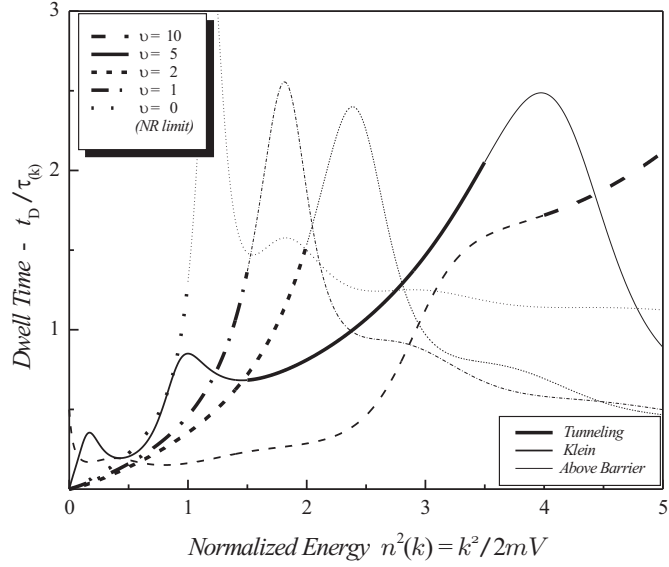,width=10cm}
\vspace{-0.6 cm}
\caption{The related tunneling PHASE times in correspondence with their tunneling TRANSMISSION probability for the dynamics of the Klein-Gordon form equation.
Again, by means of the line thickness we observe that the tunneling region is comprised by the interval $(n^{\2} - \upsilon/2)^{\2} < 1,\,\, n^{\2} > 0$.
We have used the same criteria and the same set of parameters from the Fig.~\ref{Fig01}.}
\label{Fig03}
\end{figure}

At this point, could one say metaphorically that the particle represented by the positive energy incident wave packet spend a time equal to $ t_{T, \varphi}$ inside the barrier before retracing its steps or tunneling?
The answer is in the definition of the dwell time for the relativistic colliding configuration which we have proposed.
In quantum mechanics, using steady-state wave functions, the average time of residence in a region is the integrated density divided by the total flux in (or out) and the lifetime is defined as the difference between these residence times with and without interactions.
The dwell time is a measure of the time spent by a particle in the barrier region regardless of whether it is ultimately transmitted or reflected \cite{But83}.
Therefore, in parallel to the above way of calculating the traversal time, we believe that a more complete or, at least, complementary answer to such an inquiry emerges with the definition of the dwell time for the same tunneling configuration evaluated with a the Klein-Gordon equation,
\small\begin{equation}
t_{\D}
=\frac{m}{k} \int_{\0}^{\L}{\mbox{d}x{|\phi_{\2}(k,x)|^{\2}}}
\label{008}
\end{equation}\normalsize
where  $j_{in}$ is the flux of positive energy incident particles and $\phi_{\2}(k,x)$ is the stationary state wave function inside the barrier.
The normalized dwell time is thus given by
\small\begin{equation}
\frac{t_{\D}}{\tau_{\bb{k}}} = \frac{f_{\D}\bb{n, L}}{g_{\D}\bb{n, L}}
\label{009}
\end{equation}\normalsize
where
\small\begin{eqnarray}
f_{\D}\bb{n, L} &=&\left(1 - \frac{n^{\2}}{\rho\bb{n}^{\2}}\right) + \left(1 + \frac{n^{\2}}{\rho\bb{n}^{\2}}\right) \frac{Sh(\rho\bb{n} w L)\,Ch(\rho\bb{n} w L)}{\rho\bb{n} w L},
\nonumber
\end{eqnarray}\normalsize
and
\small\begin{eqnarray}
g_{\D}\bb{n, L} &=& 2\sqrt{1 + 2 n^{\2}\upsilon} \left[1 + \frac{Sh(\rho\bb{n} w L)^{\2}}{4 n^{\2} \rho\bb{n}^{\2}}\right].
\nonumber
\end{eqnarray}\normalsize
We have computed the dwell times for the solutions of the relativistic wave equation (\ref{002}) and we have illustrated it in the Fig.~\ref{Fig03} in correspondence with the transmission probabilities of the Fig.~\ref{Fig01}.
\begin{figure}
\vspace{-0.6 cm}
\epsfig{file=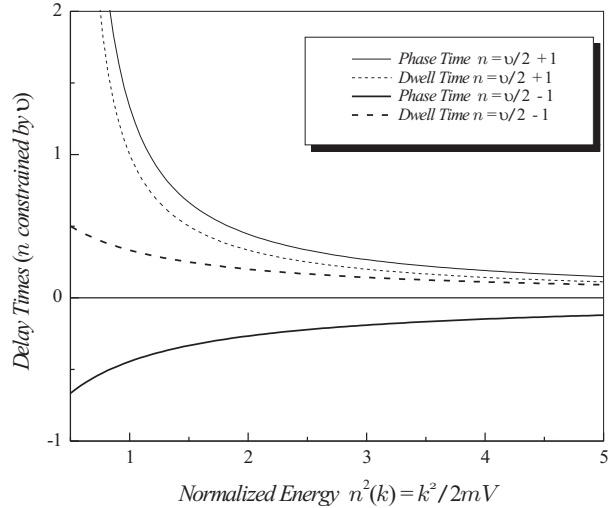,width=10cm}
\vspace{-0.6 cm}
\caption{Phase times (solid lines) and dwell times (dashed lines) in the limits of the tunneling energy zone as a function of the propagation regime (normalized energy parameter $n^{\2}\bb{k}$).
The analytical transition between the tunneling zone and the above barrier zone is given by $n^{\2}= \upsilon/2 + 1$, and between the tunneling zone and the Klein (paradox) zone by $n^{\2}= \upsilon/2 - 1$.
In this lower limit, $n^{\2}\rightarrow \upsilon/2 - 1$, the phase times result in negative values.}
\label{Fig04}
\end{figure}
By taking the same limits that we have considered for the phase time, the expressions for the dwell time provide us with,
\small\begin{equation}
\lim_{n^{\2}\rightarrow \upsilon/2 \mp 1}{\frac{t_{\D}}{\tau_{\bb{k}}}} = \frac{1}{2}\frac{1}{2 n^{\2} \pm 1}, ~~~~ n^{\2}\rightarrow \upsilon/2 \mp 1, ~~~n^{\2},\,\upsilon > 0.
\label{014}
\end{equation}\normalsize
from which we plot the comparative results in the Fig.~\ref{Fig04}.

In general lines, we see that the transmission probability depends only weakly on the barrier height, approaching the perfect transparency for very high barriers, in stark contrast to the conventional, non-relativistic tunneling where $T\bb{n, L}$ exponentially  decays with the increasing $V_{\0}$.
Such a relativistic effect is usually attributed that a sufficient strong potential, being repulsive for electrons, is attractive for positrons and results
in positron states inside the barrier, which align the energy with the electron continuum outside \cite{Kat06}.

By analyzing the magnitude of the parameter $m L$ ($(m c^{\2})/[\hbar (c/L)]$ in standard units) for an electron with mass $\sim 0.5\,MeV$, since in natural units we have $0.2\, MeV\, pm \sim 1$ it should be necessary a potential barrier of width $L << 1 \,pm$ to permit the observation of such a superluminal transmission.
In the most common sense, it is natural since we are simply saying that the Compton wavelength ($\hbar/(m c)$) is much larger than the length of the potential barrier that, in this limit situation, becomes invisible (transparent) for the tunneling particle.

\section{Conclusions}

Here we have been concentrated on the accurate calculation of phase times and dwell times for the energy zone of tunneling governed by a relativistic wave equation.
We also have quantified the conditions for the occurrence of accelerated and, eventually, superluminal tunneling transmission probabilities in order to eliminate the called filter (Hartman) effect from the transmitted waves.
By eliminating the filter effect, the transmission probabilities approximates the unitary modulus (complete transmission through a {\em transparent} medium).
As a result, the stationary phase conditions could be accurately considered for evaluating the phase time which gives the {\em exact} position of the transmitted wave packet (or particles).
It represents a novel possibility for accelerated tunneling transmission, different from the standard results of the non-relativistic analysis, for which, due to the distortion of the original momentum distribution, the position of the peak is shifted.

Concerning to the implications of the magnitude of the parameter $m L$ ($m c^{\2}/[\hbar (c/L)]$ in standard units) for the effective observation of such effects, in the most common sense, it is important to notice that such restrictive condition should be naturally expected since we are simply assuming that the Compton wavelength ($\hbar/(m c)$) is much larger than the length $L$ of the potential barrier that, in this limit situation, becomes {\em invisible} for the tunneling particle.
The relativistic quantum mechanics establishes that if a wave packet is spread out over a distance $d >> 1/m$, the contribution of momenta $|p| \sim m >> 1/d$ is heavily suppressed, and the negative energy components of the wave packet solution are negligible; the one-particle theory is then consistent.
However, if we want to localize the wave packet in a region of space (wave packet width $d$) smaller than or of the same size as the Compton wavelength, that is $d < 1/m$, the negative energy solutions (positron states) start to play an appreciable role.

If we had pursued a more detailed analysis involving fermions, the above qualitative arguments could report us to the Klein paradox and the creation of particle-antiparticles pairs during the scattering process which might create the intrinsic (polarization) mechanisms for accelerated and/or non-causal fermion teletransportation.
The Klein paradox is intrinsically related to the quoted Klein tunneling \cite{Del03,Cal99,Dom99,Che02}.

Furthermore, by analyzing the phenomenon from the Dirac equation, the condition $d < L < 1/m$ imposed over a positive energy component of the incident wave packet in the relativistic tunneling configuration excite the negative energy modes (antiparticles) in the same way that the movement of electrons in a semi-conductor is concatenated with the movement of positively charged {\em holes}.
Thus, despite the theoretical focus, the results here obtained, for the Klein-Gordon dynamics, give some indications to what configurations should deserve further attention by experimenters, for instance, the study of the graphene structures, where the dynamics of the electron is described by a relativistic Dirac-like dynamics \cite{Kat06}.

In graphene, due to the chiral nature of their quasiparticles, quantum tunneling in these materials becomes highly anisotropic, qualitatively different from the case of normal, non-relativistic electrons.
The graphene provides an effective medium for mimicking relativistic quantum effects where, for instance, massless Dirac fermions allow a close realization of Klein's gedanken experiment whereas massive chiral fermions in bilayer graphene offer an interesting complementary system that elucidates the basic physics involved.
In this scenario, we just notice that the Klein tunneling and the accelerated tunneling transmission associated with relativistic-like phenomena at nanoscopic scale deserves a more careful study, once it can eventually be tested experimentally.


{\bf Acknowledgments}
We would like to thank FAPESP (PD 04/13770-0) for the financial support.


\begin{thebibliography}{99}

\bibitem{But03}
M. B\''{u}ttiker and S. Washburn, Nature {\bf 422}, 271 (2003).
\bibitem{Win03A}
H. G. Winful, Nature {\bf 424}, 638 (2003).
\bibitem{Win03}
H. G. Winful, {\em Phys. Rev. Lett.} {\bf 91}, 260401 (2003).
\bibitem{Olk04}
V. S. Olkhovsky, E. Recami and J. Jakiel, {\em Phys. Rep.} {\bf 398}, 133 (2004).
\bibitem{Nim92}
A. Enders, G. Nimtz, J. Phys.-I (France) {\bf 2}, 1693 (1992).
\bibitem{Ste93}
A. M. Steinberg, P. G. Kwiat and R. Y. Chiao, Phys. Rev. Lett.{\bf 71}, 708 (1993).
\bibitem{Spi94}
C. Spielmann, R. Szipöcs, A. Stingl and F. Krausz, Phys. Rev. Lett. {\bf 73}, 2308 (1994).
\bibitem{Nim94}
G. Nimtz, A. Enders and H. Spieker,{\em J. Phys.-I} (France) {\bf 4}, 1 (1994).
\bibitem{Hay01}
A. Haybel and G. Nimtz, Annals Phys. {\bf 10}(Leipzig), Ed.08, 707 (2001).
\bibitem{Hau89}
E. H. Hauge and J. A. Stovneng, {\em Rev. Mod. Phys.} {\bf 61}, 917 (1989).
\bibitem{Sok87}
D. Sokolorski and L. M. Baskin, {\em Phys. Rev.} {\bf A36}, 4604 (1987).
\bibitem{Ima97}
K. Imafuku, I. Ohba, Y. Yamanaka, {\em Phys. Rev.} {\bf A56}, 1142 (1997).
\bibitem{Abo00}
M. Abolhasani and M. Golshani, {\em Phys. Rev.} {\bf A62}, 012106 (2000).
\bibitem{Bro94}
S. Brouard, R. Sala, and J. G. Muga, {\bf Phys. Rev.} {\bf A49}, 4312 (1994).
\bibitem{But83}
M. B\"{u}ttiker, {\em Phys. Rev.} {\bf B27}, 6178 (1983).
\bibitem{But82}
M. B\"{u}ttiker and R. Landauer, {\em Phys. Rev. Lett.} {\bf 49}, 1739 (1982).
\bibitem{Wig55}
E. P. Wigner, {\em Phys. Rev.} {\bf 98}, 145 (1955).
\bibitem{Lan94}
R. Landauer and Th. Martin, {\em Rev. Mod. Phys.} {\bf 66}, 217 (1994).
\bibitem{Olk92}
V. S. Olkhovsky and E. Recami, {\em Phys. Rep.} {\bf 214}, 339 (1992).
\bibitem{Olk95}
V. S. Olkhovsky, E. Recami, F. Raciti and A.K. Zaichenko, {\em J. de Physique-I} (France) {\bf 5}, 1351 (1995).
\bibitem{Jak98}
J. Jakiel, V. S. Olkhovsky, and E. Recami, {\em Phys. Lett.} {\bf A248}, 156 (1998).
\bibitem{Olk02}
V. S. Olkhovsky, E. Recami, and G. Salesi, {\em Europhys. Lett.} {\bf 57}, 879 (2002).
\bibitem{Gar70}
C. G. B. Garret and D. E. McCumber, {\em Phys. Rev.} {\bf A01}, 305 (1970).
\bibitem{Chu82}
S. Chu and W. Wong, {\em Phys. Rev. Lett.} {\bf 48}, 738 (1982);\\
B. Segard and B. Macke, {\em Phys. Lett.} {\bf A109}, 213 (1985);\\
M. W. Mitchell and R. Y. Chiao, {\em Phys. Lett.} {\bf A230}, 122 (1997);\\
L. J. Wang, A. Kuzmich and A. Dogariu, {\em Nature} {\bf 406}, 277 (2000).
\bibitem{Har62}
T. E. Hartman, {\em J. Appl. Phys.} {\bf 33}, 3427 (1962).
\bibitem{Del03}
F. Delgado {\it et al.}, Phys. Rev. {\bf A68}, 032101 (2003).
\bibitem{Cal99}
A. Calogeracos and N. Dombey, Int. J. Mod. Phys. {\bf A14}, 631 (1999).
\bibitem{Dom99}
N. Dombey and A. Calogeracos, Phys. Rep. {\bf 315}, 41 (1999).
\bibitem{Che02}
Chun-Fang Li and Xi Chen, Ann. Phys. {\bf 12} (Leipzig), Ed.10, 916 (2002).
\bibitem{Kle29}
O. Klein, Z. Phys. {\bf 53}, 157 (1929).
\bibitem{Zub80}
C. Itzykson and J. B. Zuber, {\it Quantum Field Theory} (Mc Graw-Hill Inc., New York, 1980).
\bibitem{Aux1}
R. K. Su, G. Siu and X. Chou, J. Phys. {\bf A26}, 1001 (1993);
B. R. Holstein, Am. J. Phys. {\bf 66}, 507 (1998);
H. Nitta, T. Kudo and H. Minowa, Am. J. Phys. {\bf 67}, 966 (1999).
\bibitem{Kre04}
P. Krekora, Q. Su and R. Grobe, Phys. Rev. Lett. {\bf 92}, 040406 (2004).
\bibitem{Kre01}
P. Krekora, Q. Su  and R. Grobe, Phys. Rev. {\bf A63}, 032107 (2001).
\bibitem{Pet03}
V. Petrillo and D. Janner, Phys. Rev. {\bf A67}, 012110 (2003).
\bibitem{Tel95}
V. L. Telegdi, in {\it Klein's Paradox Revisited},Ed.by U. Lingstrom (World Scientific, Singapore, 1995).
\bibitem{Han99}
A. Hansen and F. Ravndal, {\em Phys. Scr.} {\bf 23} 1036 (1999).
\bibitem{Lan89}
R. Landauer, {\em Nature} {\bf 341}, 567 (1989).
\bibitem{Ber06}
A. E. Bernardini, {\em Phys. Rev.} {\bf A74}, 062111 (2006);
A. E. Bernardini, S. De Leo and P. P. Rotelli, {\em Mod. Phys. Lett} {\bf A19}, 2717 (2004).
\bibitem{Gav84}
B. Gaveau {\it et al.}, {\em Phys. Rev. Lett.} {\bf 53}, 419 (1984).
\bibitem{Gre85}
W. Greiner, B. Mueller, and J. Rafelski, {\em Quantum Electrodynamics of Strong Fields} (Springer, Berlin, 1985).
\bibitem{Pag05}
D. N. Page, New J. Phys. {\bf 7}, 203 (2005).
\bibitem{Kat06}
M. I. Katsnelson, K. S. Novoselov and A. K. Geim, Nature Phys. {\bf 02}, 620 (2006)

\end{thebibliography}
\end{document}